\documentstyle[sprocl,epsf]{article}

\begin{document}
\title{Finite energy chiral sum rules in QCD\footnote{\*Work supported
in part by the Volkswagen Foundation}}

\author{C. A. Dominguez}
\address{Institute of Theoretical Physics and Astrophysics\\
University of Cape Town, Rondebosch 7700, South Africa}
\author{K. Schilcher}
\address{
Institut f\"{u}r Physik, Johannes Gutenberg-Universit\"{a}t\\
Staudingerweg 7, D-55099 Mainz, Germany}
\maketitle
\begin{abstract}
\abstracts{
A set of well known chiral sum rules, expected to be valid in QCD, is
confronted with experimental data on the vector and axial-vector
hadronic spectral functions, obtained from tau-lepton decay by the ALEPH
collaboration. The Das-Mathur-Okubo sum rule, the first and second
Weinberg sum rules, and the electromagnetic pion mass difference sum
rule are not well saturated by the data. Instead, a modified set of
sum rules having additional weight factors that vanish at the end
of the integration range on the real axis, is found to be precociously
saturated by the data to a remarkable extent.
}
\end{abstract}

There is a set of sum rules, first discovered in the framework 
of current algebra, which are now understood as consequences 
of the underlying chiral symmetry of QCD. We refer to the Das-Mathur-Okubo
sum rule (DMO) \cite{DMO}
\begin{eqnarray*}
W_{0}\equiv\int_{0}^{\infty} \frac{ds}{s}
\left[\rho_{V} (s)-\rho_{A} (s) \right]
   = \frac{1}{3} f_{\pi}^{2} <r_{\pi}^{2}>-F_{A}
\end{eqnarray*}
\begin{equation}
= - 4 \bar{L}_{10}=(2.73 \pm 0.12) \times 10^{-2}\; ,
\end{equation}
the first and second Weinberg sum rules \cite{WSR}
\begin{equation}
W_{1} \equiv
\int_{0}^{\infty} \; ds \;
\left[ \rho_{V} (s) - \rho_{A} (s) \right] = f_{\pi}^{2} \; ,
\end{equation}
\begin{equation}
W_{2} \equiv
\int_{0}^{\infty} \; ds \; s \;
\left[ \rho_{V} (s) - \rho_{A} (s) \right] = 0 \; ,
\end{equation}
and the electromagnetic pion mass difference sum rule \cite{PMD}
\begin{equation}
W_{3} \equiv
\int_{0}^{\infty} \; ds \; s \;{\rm ln} \;
\left( \frac{s}{\mu^{2}} \right) \;
\left[ \rho_{V} (s) - \rho_{A} (s) \right]
= - \frac{16 \; \pi^{2} \; f_{\pi}^{2}}{3 \; e^{2}} \;
\left( \mu_{\pi^{\pm}}^{2} \; - \mu_{\pi^{0}}^{2} \right) \; . 
\end{equation}
In the above equations $\rho_{V,A}(s)$ are, respectively, the imaginary 
parts of the vector and axial-vector two-point functions
\begin{equation}
 \Pi_{\mu \nu}^{VV} (q^2) = i \int  d^4  x  e^{i q x}
  <0|T(V_{\mu}(x) V_{\nu}^{\dagger}(0))|0> 
  = (- g_{\mu \nu} q^{2} + q_{\mu} q_{\nu}) \Pi_{V} (q^{2}) \; ,
\end{equation}
\begin{eqnarray*}
 \Pi_{\mu \nu}^{AA} (q^2) = i \; \int \; d^4 \; x \; e^{i q x} \; \;
  <0|T(A_{\mu}(x) \; \; A_{\nu}^{\dagger} (0) )|0> \; 
\end{eqnarray*}
\begin{equation}
  = \; (- g_{\mu \nu} \; q^{2} + q_{\mu} q_{\nu}) \; \Pi_{A} (q^{2})
  - q_{\mu}  q_{\nu} \; \Pi_{0} (q^{2})  \; ,
\end{equation}
where $V_{\mu}(x) = :\bar{q}(x)  \gamma_{\mu}  q(x):$,
$A_{\mu}(x) = :\bar{q}(x)  \gamma_{\mu}  \gamma_{5} q(x):$, with
$q=(u,d)$, the pion decay constant is $f_{\pi} = 92.4 \pm 0.26 \; 
\mbox{MeV}$ \cite{PDG}, the pion electromagnetic mean square radius
has the value $<r^{2}_{\pi}> = 0.439 \pm 0.008 \; \mbox{fm}^{2}$ \cite{AMEN},
and $F_{A}$ is the axial-vector coupling measured in radiative pion decay,
$F_{A} = 0.0058 \pm 0.0008$ \cite{PDG}. 
The above set of sum rules are exact relations in QCD, in
the chiral $SU(2)_{L} \times SU(2)_{R}$ limit, i.e. for vanishing up- 
and down-quark masses \cite{RA}. At short distances, the spectral 
functions 
are strictly identical in the chiral limit ($m_{q}=0$), to all orders in 
QCD perturbation theory, and up to dimension-four non-perturbatively.   
The difference between $\rho_{V}$ and  $\rho_{A}$, always at short 
distances, is non-perturbative and starts at dimension-six involving 
the four-quark condensate.  On the other hand,
the low energy behaviour of $\Pi_{V-A} (q^{2})$ is governed by chiral 
perturbation theory. In confronting these sum rules with experimental
data on the spectral functions, one must bear in mind that the latter
are known up to a certain finite energy $\sqrt{s_{0}}$; in the case of 
the ALEPH data \cite{ALEPH} this is $s_{0} < M_{\tau}^{2} < \infty $. 
Invoking quark-hadron duality, the sum rules Eqs. (1)-(4) become effectively
Finite Energy Sum Rules (FESR). The question is then, how well is local
and global duality satisfied by the experimental data. It has been argued
recently \cite{CADKS} that
the agreement between data and theory, which measures
the validity of local duality, is not entirely satisfactory. 
In view of this, one would expect a similar unsatisfactory saturation of 
the four chiral sum rules, Eqs.(1)-(4), and  in fact, this is what we find. 
Before illustrating our results, let us write down a set of four chiral
sum rules modified by an integral kernel vanishing at $s=s_{0}$ on the
real axis. This modification may be  referred to as {\it restricted global 
duality}, which is expected to set in much sooner than ordinary global
duality. The sum rules are
\begin{equation}
\bar{W}_{0} \;  \equiv \; \int_{0}^{\infty} \; \frac{ds}{s} \;
(1 - \frac{s}{s_{0}})\;
\left[ \rho_{V} (s) - \rho_{A} (s) \right]
   =- 4 \;\bar{L}_{10} \;- \; \frac{f_{\pi}^{2}}{s_{0}}    \; ,
\end{equation}
\begin{equation}
\bar{W}_{1} \equiv
\int_{0}^{\infty} \; ds \;  (1- \frac{s}{s_{0}})\;
\left[ \rho_{V} (s) - \rho_{A} (s) \right] 
= f_{\pi}^{2} \; ,
\end{equation}
\begin{equation}
\bar{W}_{2} \equiv
\int_{0}^{\infty} \; \frac{ds}{s} \; (1- \frac{s}{s_{0}})^{2}\;
\left[ \rho_{V} (s) - \rho_{A} (s) \right] \;  = \;
- 4 \bar{L}_{10} - 2 \frac{f_{\pi}^{2}}{s_{0}} \; ,
\end{equation}
\begin{equation}
\bar{W}_{3} \equiv
\int_{0}^{\infty} \; ds \; s \;{\rm ln} \;
\left( \frac{s}{s_{0}} \right) \;
\left[ \rho_{V} (s) - \rho_{A} (s) \right]
= - \frac{16 \; \pi^{2} \; f_{\pi}^{2}}{3 \; e^{2}} \;
\left( \mu_{\pi^{\pm}}^{2} \; - \mu_{\pi^{0}}^{2} \right) \; ,
\end{equation}
where Eq.(7) is a combination of the DMO sum rule and the first 
Weinberg sum rule, Eq.(9) is a combination of the DMO sum rule and
the first and second Weinberg sum rules,
and the arbitrary scale in Eq.(10) has been fixed
to $s_{0}$, which becomes the upper limit of integration in the four sum
rules. We now show that these modified sum rules are 
saturated far better than the original sum rules Eqs.(1)-(4).
In Fig.1 we plot the left hand side (l.h.s.) of
Eq.(1) computed using the fit to the data (curve(a)), and the right hand
side (r.h.s.) (curve(b)). Agreement with the data can be considerably
improved by rescaling the r.h.s. of Eq.(1) from the value $2.73 \times
10^{-2}$ to $ 2.43 \times 10^{-2}$. Figure 2 shows the l.h.s. of the 
modified DMO sum rule Eq.(7) (curve(a)) compared to the r.h.s. (curve (b))
after performing the above rescaling. 
In Fig.3 we plot the l.h.s. of the 
first Weinberg sum rule Eq.(2) (curve(a)), its modified version, 
Eq.(8) (curve(b)), and their r.h.s. (curve(c)). Figure 4 shows 
the l.h.s. of the second Weinberg sum rule Eq.(3) (curve(a)), 
compared to its r.h.s. (curve(b)), and Fig. 5 the corresponding curves 
for the modified sum rule,
Eq.(9). Finally, in Fig. 6 we plot the l.h.s. of the sum rule Eq.(4)
(curve(a)), the l.h.s. of the modified sum rule, Eq.(10) (curve (b)), and
their r.h.s. (curve(c)). An inspection of these results clearly indicates
that, the original
chiral sum rules do not appear well saturated by the data. While an
overall constant rescaling of the experimental data can result in a
better saturation of the DMO sum rule, this would not help with the other
three sum rules. Hence, the problem cannot be blamed on a systematic
overall normalization uncertainty in the data. On the other hand, by
using {\it restricted global duality}, the four modified chiral sum
rules Eqs.(7)-(10) are extremely well saturated by the data.

\begin{figure}[tb]
\epsfxsize=10cm
\epsfysize=7cm
\epsffile{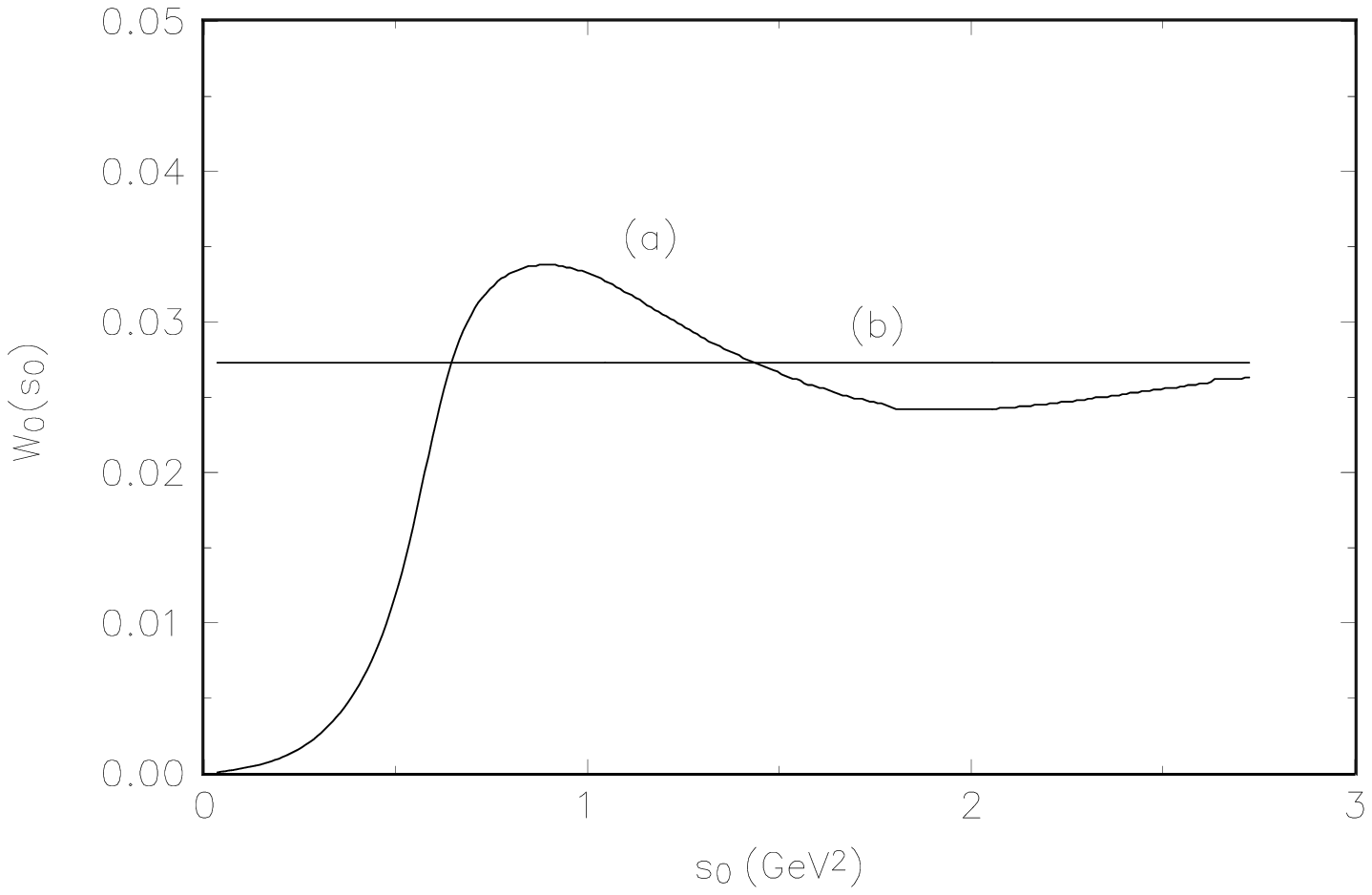}
\caption{}
\end{figure}

\begin{figure}[tb]
\epsfxsize=10cm
\epsfysize=7cm
\epsffile{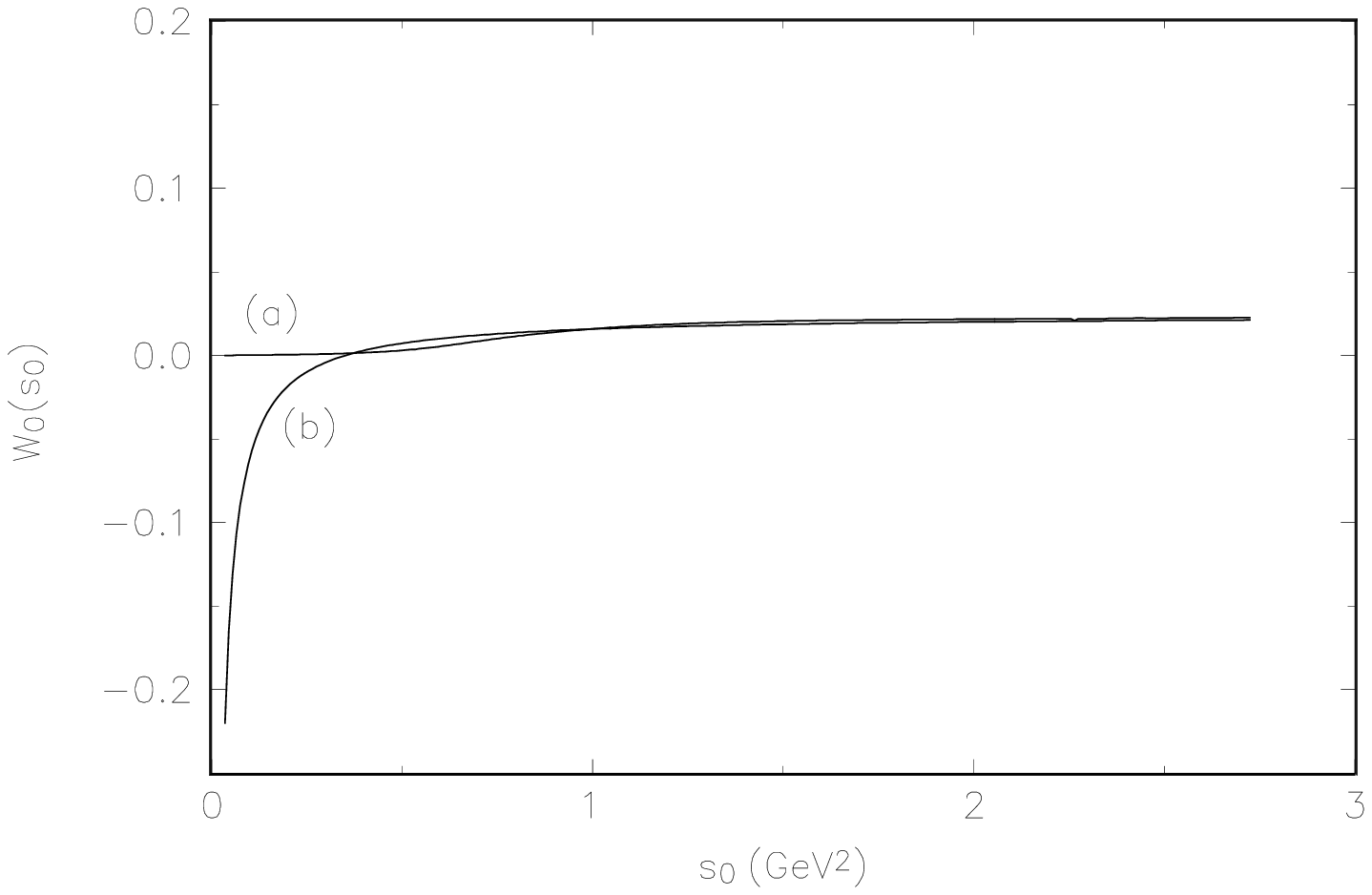}
\caption{}
\end{figure}

\begin{figure}[tb]
\epsfxsize=10cm
\epsfysize=7cm
\epsffile{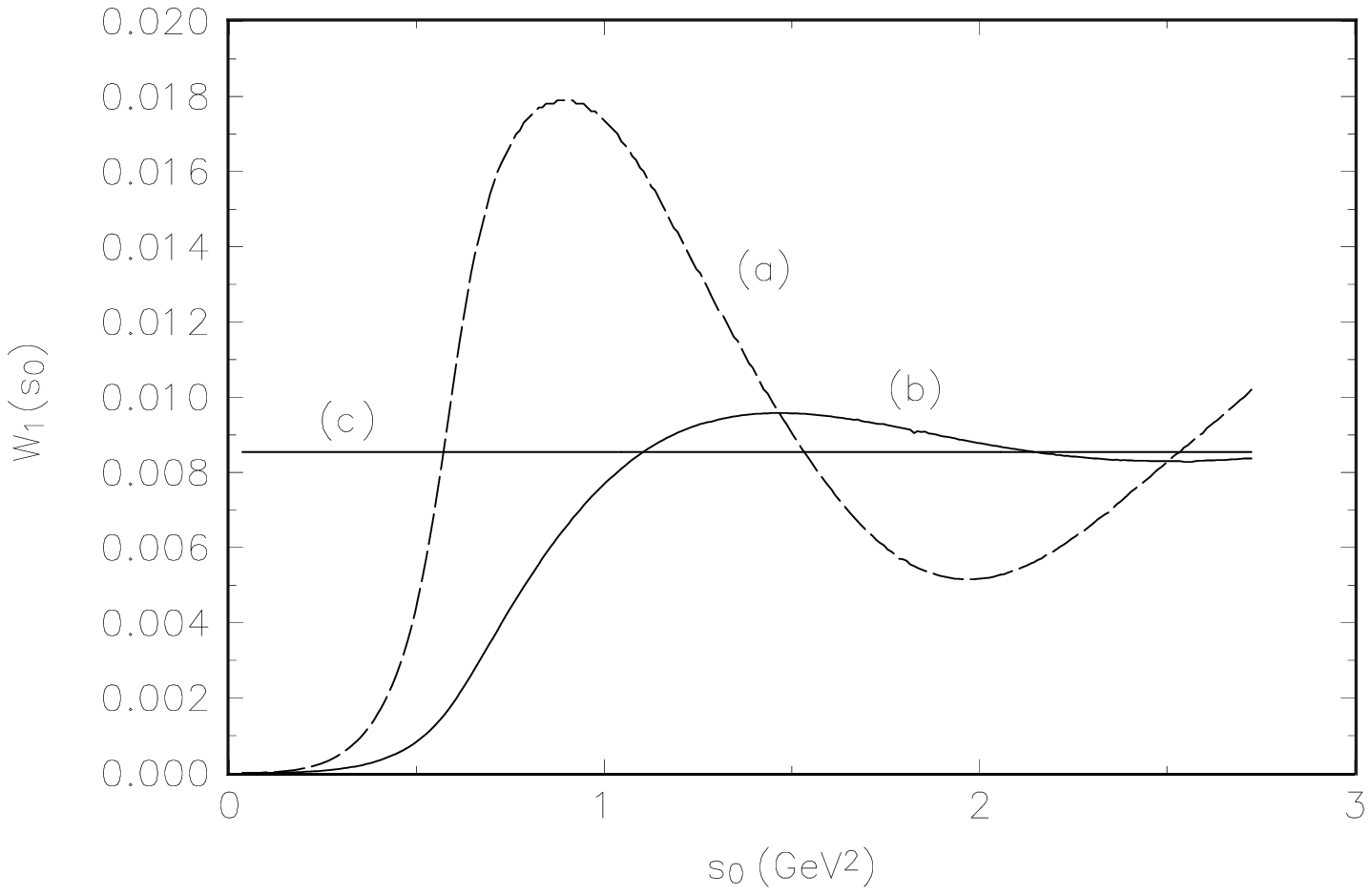}
\caption{}
\end{figure}

\begin{figure}[tb]
\epsfxsize=10cm
\epsfysize=7cm
\epsffile{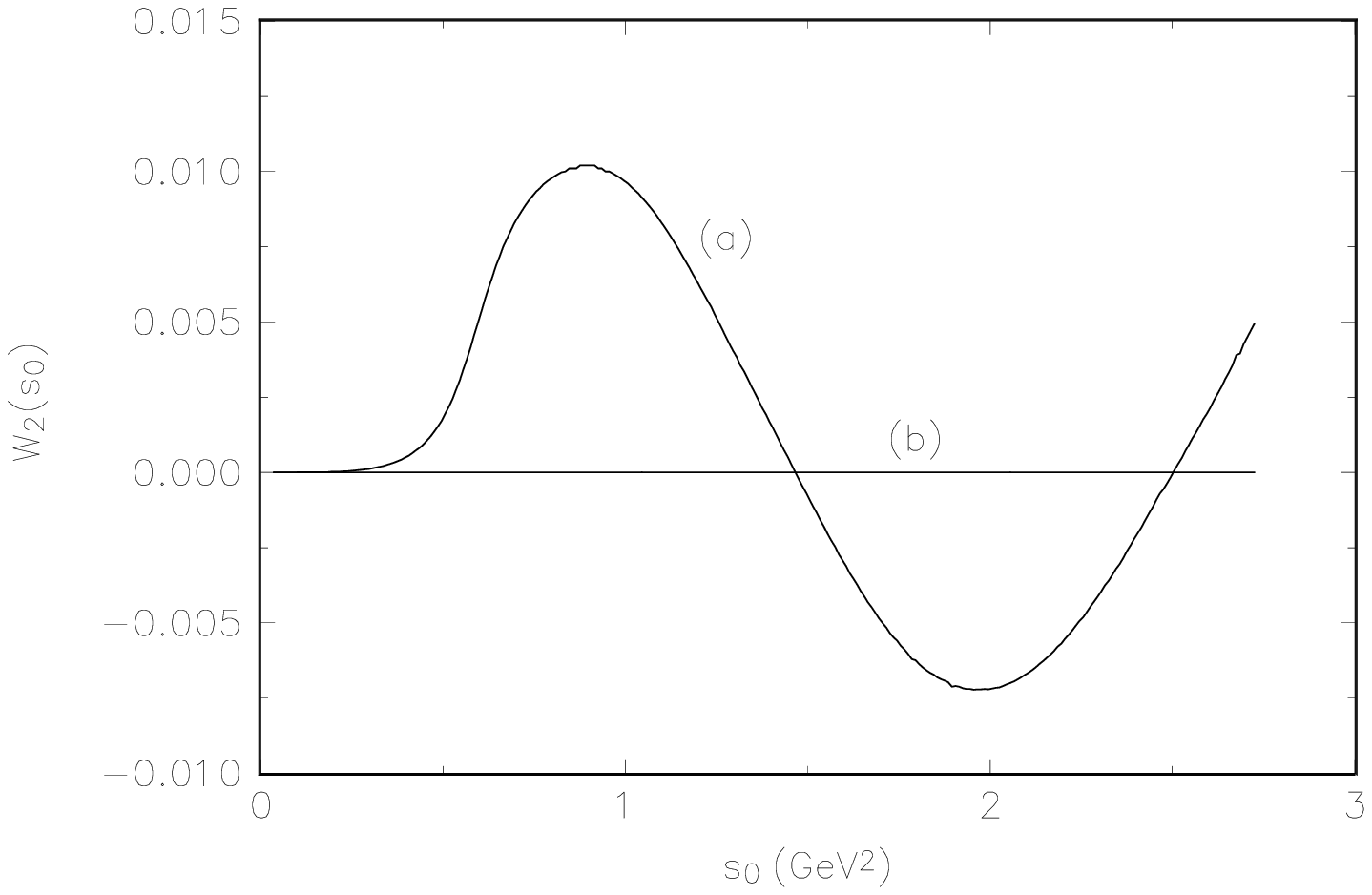}
\caption{}
\end{figure}

\begin{figure}[tb]
\epsfxsize=10cm
\epsfysize=7cm
\epsffile{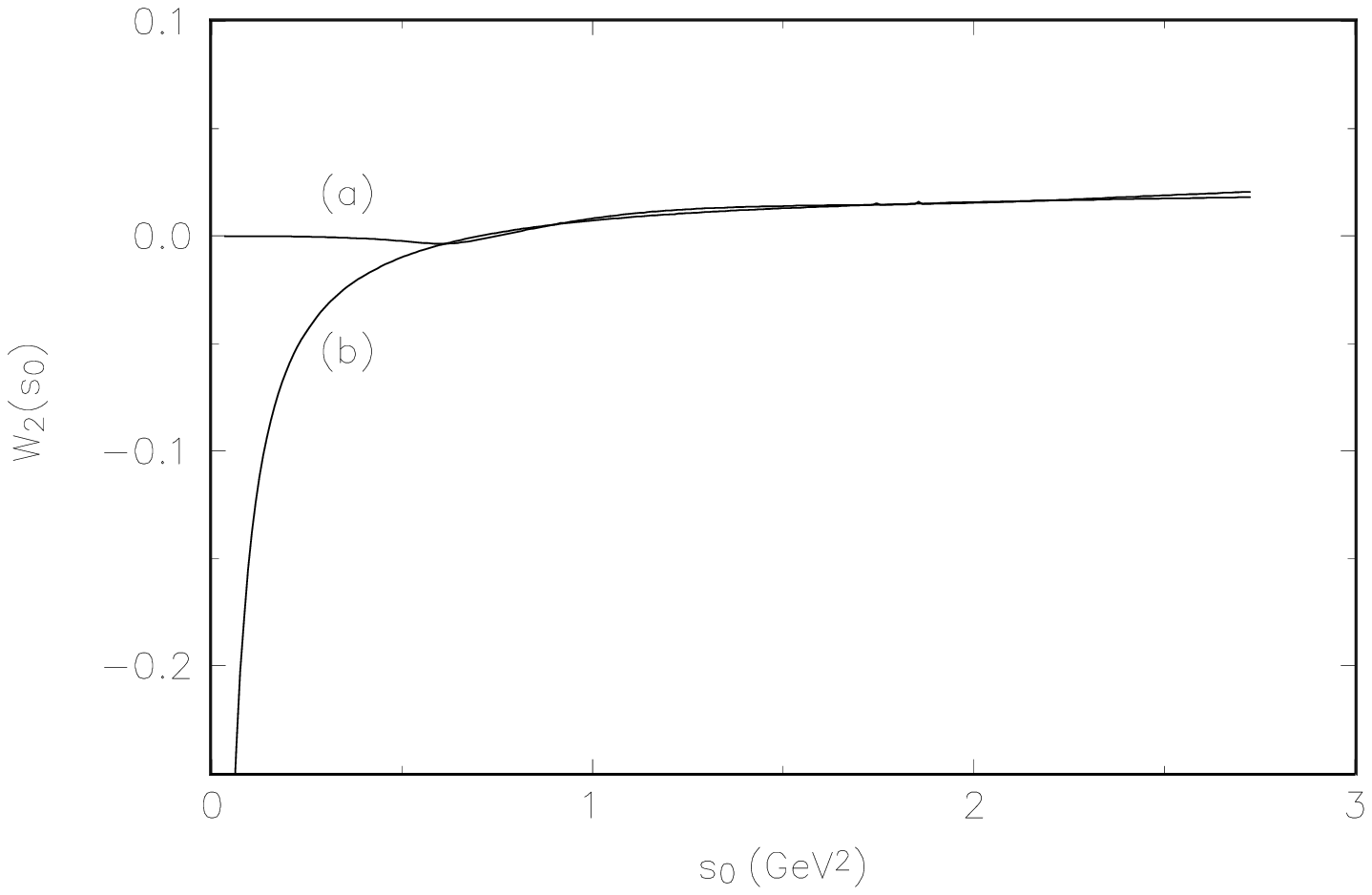}
\caption{}
\end{figure}

\begin{figure}[tb]
\epsfxsize=10cm
\epsfysize=7cm
\epsffile{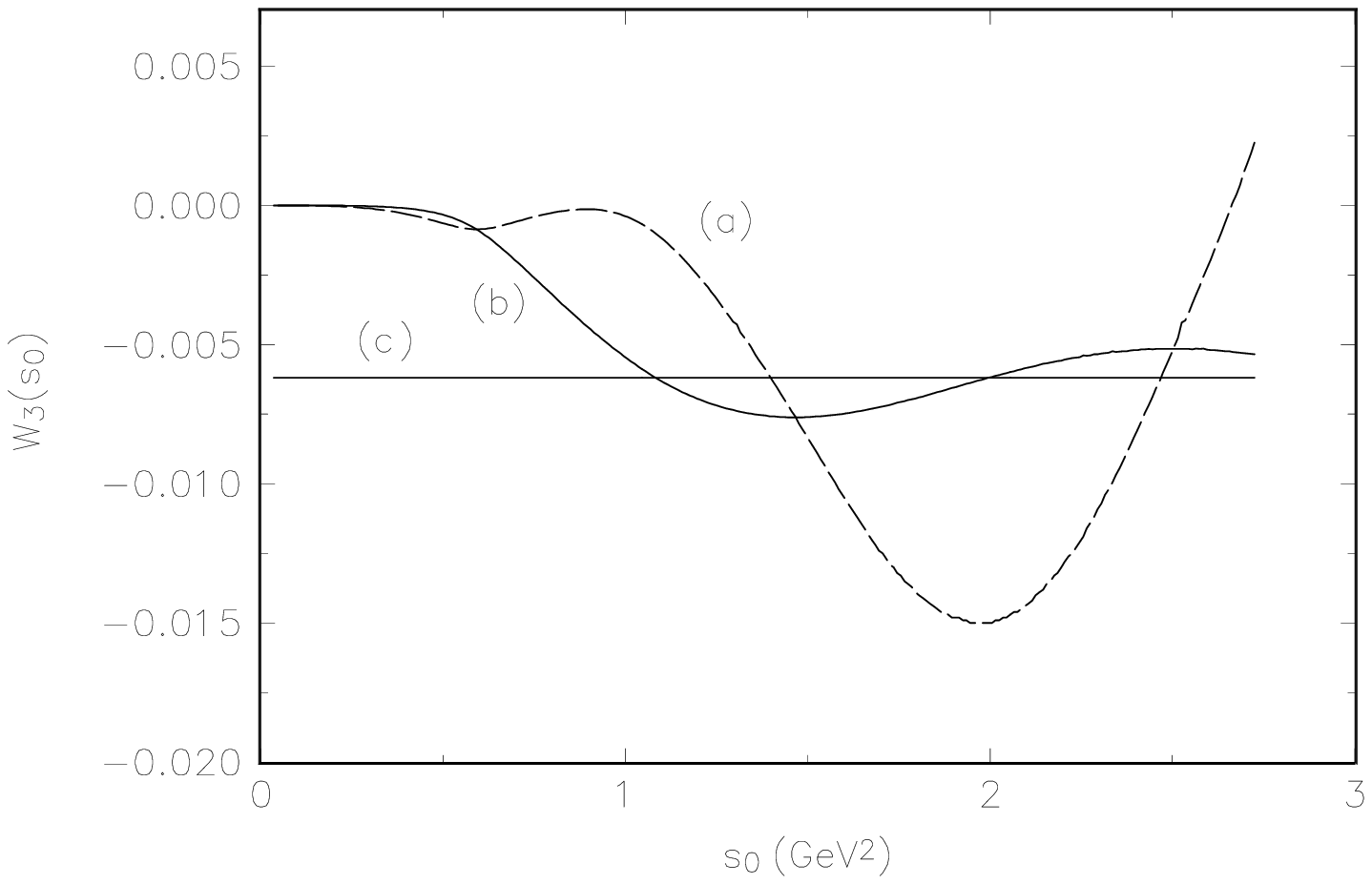}
\caption{}
\end{figure}

\end{document}